# Hidden variable models for entanglements can or cannot have a local component?

Sofia Wechsler [1)]

**Abstract**

A recent article of Colbeck and Renner tackled the problem whether entanglements may be explained by combined models of local and non-local hidden variables. To the difference from previous works they considered models in which each pair of entangled particles behaves in the same way, and the particles in the pair are equivalent, i.e. each of them produces its response to a measurement according to both local and non-local hidden variables. Their article aimed at proving that the local hidden variable component in such models has no effect on the measurement results, i.e. only the non-local variables are relevant. However, their proof deals with a very restrictive case and assumes questionable constraints on the hidden variables.

The present text studies the Colbeck and Renner class of models on a less restrictive case and under no constraints on the hidden variables. It is shown again that the local component cannot have any influence on the results. However, the Colbeck and Renner class of models is not the only one possible. A different class is described, which does admit local hidden variables by the side of the non-local influence. This class presents a couple of advantages.

## 1. Introduction

After the failure of the models of purely local hidden variables to explain the correlations appearing in quantum entanglements, different authors proposed combined models of hidden variables (HV). Elitzur, Popescu, and Rohrlich [1], then Barrett, Kent, and Pironio [2], showed that in maximally entangles states of two quantum systems, if one writes the quantum correlations as a mixture of local correlations and general (not necessarily quantum) correlations, the coefficient of the local correlations must be zero.

In a recent article, Colbeck and Renner (henceforth C&R) examined another class of local and non-local models in which all the pairs of particles are equivalent, and in each pair the response of each particle is determined by both the local and the non-local HV [3]. They showed that for these models the local HV component cannot have any effect on the measurement results.[2)]

However, their proof is based on a very restricted case, a pair of photons in polarization-singlet with the two photons tested along almost the same direction in the space. Also, they assumed that both the local and non-local HV obey classical distributions of probabilities, assumption particularly questionable because it is not clear what is a non-local HV, no description of such a thing is given.

In the present article, the Colbeck and Renner class of models is analyzed on a more general state, see section **2**, and without probabilities. Again the conclusion emerges, section **3**, that the local HV component cannot have any effect on the measurement results. Neither the state examined here is the most general type of entanglement. Though, it is plausible that the nature is consistent with itself. Indeed, there are three major problems that the non-local influence has to solve, and they are listed below. If for some entangled states the nature is able to solve these problems without local HV, then it is plausible that so it does with all the entanglements. And here are those major problems:

A) No particle may travel at superluminal velocity to serve as a carrier for the non-local influence.

---

[1)] Computers Engineering Center, Nahariya, P.O.B. 2004, 22265, Israel
[2)] In fact, the Colbeck and Renner article was meant to prove the invalidity of such models at a general level. Such a model, proposed by Leggett [4], predicted a conflict with the quantum theory, and experiments were performed to test it [5, 6]. The experiments confirmed the predictions of the quantum theory.

B) In order to be able to transmit information to one another, two parties have to know one another, which doesn't happen in case the entangled particles never met, see the experiment in the next section and [7].

C) The correlations lead to deadlock: taking as an example two entangled particles, *1* and *2*, the result of the test on particle *1* depends on the result of the test on particle *2*; but the result of the test on particle *2* also depends on the result of the test on particle *1*. The pair should be stuck and produce no results at all. But in our experiments we get results.

A. Suarez and his co-workers [8] proposed for the problem C) a solution in which the symmetry assumed in [3] in the behavior of the particles, is broken. In the Suarez' model, a) one particle in the pair produces its result independently, and the other particle produces its result dependently, as enforced by the correlations and the result of the independent particle; b) the criterion for independent behavior, is the order of the events of impingement on certain choice-devices: the particle first to impinge on its choice-device is independent. However the experiments [9, 10] disproved this model.

It is not clear whether both Suarez' assumptions are wrong, or only one of them and in this case which one. The present article also examines the idea of asymmetry in the behavior of the entangled particles, but in a different way with respect to the point b), and that presents some advantages.

## 2. An entanglement of 3 particles from different sources

Consider three identical sources of identically polarized fermions $S_A$, $S_B$, $S_C$. Assume that the beams are of such a low intensity that individual particles may be counted, fig. 1. Since the fermions are identically polarized it won't be possible to get two fermions at once from one and the same source.

The fermions meet the beam-splitters $BS_A$, $BS_B$, $BS_C$, that produce the transformations

(1) $|A> \rightarrow 2^{-½}(|a> + i|a'>)$,  $|B> \rightarrow 2^{-½}(|b> + i|b'>)$,  $|C> \rightarrow 2^{-½}(|c> + i|c'>)$.

Then the particles meet the phase shifts α, β, γ, and the following system-state is obtained

(2) $|\psi> = 2^{-3/2}(|a> + ie^{i\alpha}|a'>)(|b> + ie^{i\beta}|b'>)(|c> + ie^{i\gamma}|c'>)$.

Opening the parentheses one gets

(3) $|\psi> = 2^{-3/2}\{|a>|b>|c> + ie^{i\gamma}|a>|b>|c'> + ie^{i\beta}|a>|b'>|c> + ie^{i\alpha}|a'>|b>|c>$

$- e^{i(\beta+\gamma)}|a>|b'>|c'> - e^{i(\alpha+\gamma)}|a'>|b>|c'> - e^{i(\alpha+\beta)}|a'>|b'>|c> - ie^{i(\alpha+\beta+\gamma)}|a'>|b'>|c'>\}$.

The particles impinge further on the beam-splitters $BS_D$, $BS_E$, $BS_F$. One can check that all the terms in (3) except the first and the last, comprise two particles landing on the same beam-splitter, and therefore exiting the same beam-splitter. Such cases may be easily identified and discarded. We are interested here in the remaining expression, so let's renormalize it

(4) $|\psi> = 2^{-3/2}\{|a>|b>|c> + e^{i\vartheta}|a'>|b'>|c'>\}$,  $\vartheta = \alpha + \beta + \gamma - \pi/2$.

The beam-splitters $BS_D$, $BS_E$, $BS_F$, produce the transformations

(5) $|a> \rightarrow 2^{-½}(|d> + i|d'>)$,  $|b> \rightarrow 2^{-½}(|e> + i|e'>)$,  $|a> \rightarrow 2^{-½}(|f> + i|f'>)$,

$|a'> \rightarrow 2^{-½}(i|e> + |e'>)$,  $|b'> \rightarrow 2^{-½}(i|f> + |f'>)$,  $|c'> \rightarrow 2^{-½}(i|d> + i|d'>)$,

which introduced in (4) yield

(6) $|\psi> = ⅛\{(|d> + i|d'>)(|e> + i|e'>)(|f> + i|f'>) + e^{i\vartheta}(i|e> + |e'>)(i|f> + i|f'>)(i|d> + i|d'>)\}$.

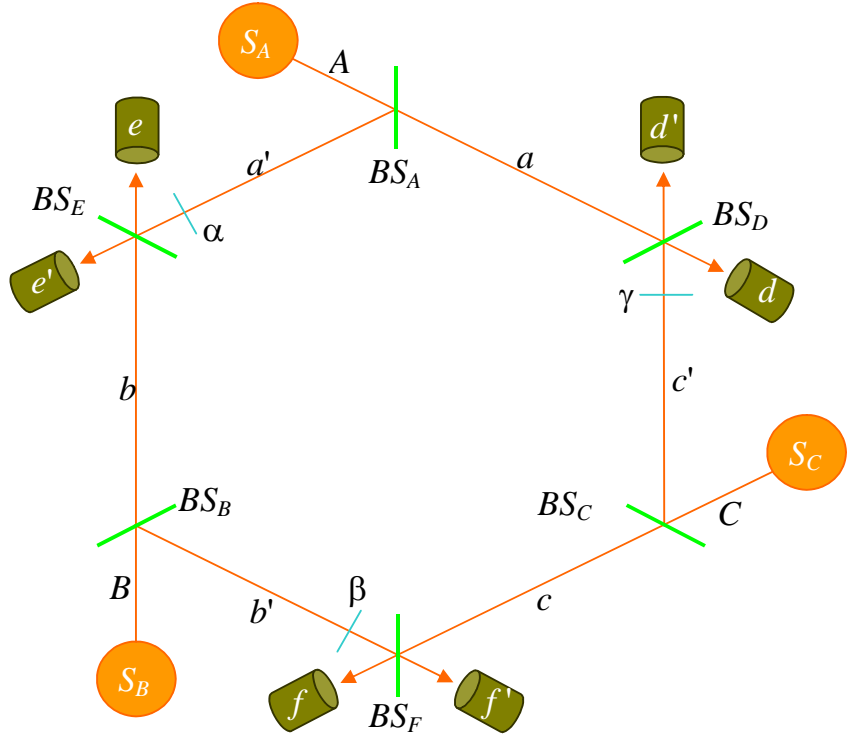

**Figure 1.** Generating an entanglement with three particles from different sources.

One can see that the two products in the R.H.S. of (6) are similar up to a permutation of the particles. Permuting two identical fermions changes the sign of the product. We have to do that twice,

$$|\psi\rangle = \tfrac{1}{8}\{(|d\rangle + i|d'\rangle)(|e\rangle + i|e'\rangle)(|f\rangle + i|f'\rangle) - e^{i\vartheta}(i|e\rangle + |e'\rangle)(i|d\rangle + i|d'\rangle)(i|f\rangle + i|f'\rangle)\}.$$

(7) $|\psi\rangle = \tfrac{1}{8}\{(|d\rangle + i|d'\rangle)(|e\rangle + i|e'\rangle)(|f\rangle + i|f'\rangle) + e^{i\vartheta}(i|d\rangle + i|d'\rangle)(i|e\rangle + |e'\rangle)(i|f\rangle + i|f'\rangle)\}.$

Now, we can open the parentheses and do summation over identical terms. For $\vartheta = \pi/2$ one gets

(8) $|\psi\rangle = \tfrac{1}{2}\{|d\rangle|e\rangle|f\rangle - |d\rangle|e'\rangle|f'\rangle - |d'\rangle|e\rangle|f'\rangle - |d'\rangle|e'\rangle|f\rangle\},$

and for $\vartheta = 3\pi/2$

(9) $|\psi\rangle = (-i/2)\{|d'\rangle|e'\rangle|f'\rangle - |d'\rangle|e\rangle|f\rangle - |d\rangle|e'\rangle|f\rangle - |d\rangle|e\rangle|f'\rangle\}.$

We denote below the detection results by $R_D$, $R_E$, and $R_F$. Each result bears the index of the beam-splitter at whose output is detected. If the detector on the primed output clicked the result is assigned the value −1, and if the detector on the unprimed output clicked the result is assigned the value +1; e.g. $R_D = -1$ if the detector $d'$ clicked and $R_D = 1$ if the detector $d$ clicked. The combinations obtained for $\vartheta = \pi/2$ have the property

(10) $R_D R_E R_F = 1,$

and the combinations obtained for $\vartheta = 3\pi/2$ have the property

(11) $R_D R_E R_F = -1.$

## 3. A symmetrical model of local and non-local HV

This section deals with the C&R class of models for entanglements. In these models each result $R_i$ is dependent on a local HV denoted here by $\lambda_i$, on a non-local HV denoted here by $\mu$, and on the set $\{\varphi\}$ of macroscopic parameters that appear in the wave function

$$(12) \quad R_i = F_i(\lambda_i, \{\varphi\}, \mu).$$

The concept of non-local HV is not clear. There is no description in the literature what is such a variable. In this text, $\mu$ expresses the non-local influence on the entangled particles, with no further assumptions and without assuming any probability distributions.

**Theorem:** *In the model (12) for the experiment described by the wave function (8) or (9), the local HVs have no effect on the experiment results.*

**Proof:** Let's begin with the wave function (8). The set $\{\varphi\}$ is the phase-shift triplet $\alpha, \beta, \gamma$, fig. 1. Let $u$, $v$, $w$, be specific values for $\lambda_D, \lambda_E, \lambda_F$, respectively. According to (10)

$$(13) \quad F_D(u, \{\varphi\}, \mu) \, F_E(v, \{\varphi\}, \mu) \, F_F(w, \{\varphi\}, \mu) = 1.$$

If $\lambda_E$ has an influence on the result $R_E$, then there has to exist also a value $u'$ of $\lambda_E$ s.t.

$$(14) \quad F_D(u', \{\varphi\}, \mu) = -F_D(u, \{\varphi\}, \mu).$$

Since the particles originate in different sources and never meet, the local hidden variables $\lambda_D, \lambda_E, \lambda_F$, are independent of one another. A change in the value of $\lambda_D$ has no influence on $\lambda_E$ and $\lambda_F$, therefore $R_E$ and $R_F$ remain the same if $u$ is replaced by $u'$. One will get in this case

$$(15) \quad F_D(u', \{\varphi\}, \mu) \, F_E(v, \{\varphi\}, \mu) \, F_F(w, \{\varphi\}, \mu) = -1$$

in contradiction with (10). Therefore no value $u'$ may exist so as to satisfy (14) if the value $u$ satisfies (13). In consequence $\lambda_D$ may have no influence on $R_D$.
The rationale continues in the same way for $\lambda_E$ and $\lambda_F$, leading to

$$(16) \quad R_i = G_i(\{\varphi\}, \mu), \quad i = D, E, F.$$

It is also obvious that the proof goes similarly for $\vartheta = 3\pi/2$. Moreover, if the state (2) may be obtained with $N$ independent particles then (10) and (11) will refer to $N$ particles and so the result (16).

## 4. An asymmetrical model of local and non-local HV

The above theorem doesn't rule out completely the possibility of local HV, because other models than (12) may be built. Let's notice that (10) and (11) show that out of the three entangled particles, two may produce arbitrary results, and only one particle has to answer dependently on the others. That suggests a model in which different particles in the entanglement behave differently.
Indeed, the nature doesn't always proceed identically with identical particles. For instance if in fig. 1 $\alpha = \beta = \gamma = \pi/3$, it may happen that two particles will give the same response, +1, and one particle the response −1. Despite the fact that the particles are identical and the setup at each station is the same, the responses are not always identical.
This situation suggests an entanglement model called below *asymmetrical*: in an experiment described by the wave function (8) or (9), two particles respond independently, eventually according to local HV at the detection stations, and one particle responds in consequence of the rest of the answers,

$$(17) \quad R_i = H_i(\lambda_i, \varphi_i), \quad R_j = H_j(\lambda_j, \varphi_j), \quad R_k = (-1)^q R_i R_j,$$

where q = 0 for $\vartheta = \pi/2$ and q=1 for $\vartheta = 3\pi/2$.

The particles that produce independent results cannot be established at the preparation step, but at the measurement step. One can see that in the two terms appearing in (4), terms obtained at the preparation step, only one particle may respond independently out of three, while according to the wave functions on which the measurement is done, (8) or (9), two particles may respond independently.

This model presents a couple of advantages: it solves the problem of the deadlock, the issue C) in section **1**, and does not need the nebulous concept of non-local HV. Still, the difficulties A) and B) are not solved.

The disadvantage is that it requires a criterion by which the independent particles are chosen. However, assume that what establishes the results $R_i$ and $R_j$ is not local parameters but "God plays dice", and that the non-local influence enforces the remaining result, $R_k$, to what the correlations demand. In this case, it's easy to show that one may regard as well $R_i$ and $R_k$, or $R_j$ and $R_k$, as being established by "God plays dice" and the remaining result as enforced by the correlations. No particle is absolutely independent.

## References

[1] A. Elitzur, S. Popescu, D. Rohrlich, "Quantum nonlocality for each pair in an ensemble", Phys. Lett. A **162**, 25, (1992).

[2] J. Barrett, A. Kent, S. Pironio, "Maximally Non-Local and Monogamous Quantum Correlations", Phys. Rev. Lett. **97**, 170409, (2006).

[3] R. Colbeck and R. Renner, "Hidden variable models for quantum theory cannot have any local part", quant -ph/0801.2218v2.

[4] A. J. Leggett, "Nonlocal Hidden-Variable Theories and Quantum Mechanics: An Incompatibility Theorem", Foundations of Phys., **33**, No. 10, page 1469, (2003).

[5] C. Branciard, A. Ling, N. Gisin, C. Kurtsiefer, A. Lamas-Linares, V. Scarani, "Experimental Falsification of Leggett's Non-Local Variable Model", Phys. Rev. Lett. **99**, 210407, (2007).

[6] S. Gröblacher, T. Paterek, R. Kaltenback, Č. Brukner, M. Żukowsky, M. Aspelmeyer, and A. Zeilinger, "An experimental test of non-local realism", Nature (London) **446**, page 871, (2007); T. Paterek, A. Fedrizzi, S. Groeblacher, T. Jennewein, M. Zukowski, M. Aspelmeyer, and A. Zeilinger, "Experimental test of nonlocal realistic theories without the rotational symmetry assumption", Phys. Rev. Lett. **99**, 210406 (2007).

[7] S. Wechsler, "A proposal of implementing Hardy's thought-experiment with particles that never met", quant-ph/0812.2582; B. Yurke and D. Stoler, "Einstein-Podolsky-Rosen Effects from Independent Particle Sources", Phys. Rev. Lett. **68**, page 1251, (1992).

[8] A. Suarez and V. Scarani, "Does entanglement depend on the timing of the impacts at the beam-splitters?", quant-ph/9704038, (1997).

[9] A. Stefanov, H. Zbinden, N. Gisin and A. Suarez, "Quantum Correlations with Spacelike Separated Beam Splitters in Motion: Experimental Test of Multisimultaneity", Phys. Rev. Lett. **88**, no. 12, (2002).

[10] N. Gisin, V. Scarani, W. Tittel and H. Zbinden, "Optical tests of quantum nonlocality: from EPR-Bell tests towards experiments with moving observers", quant-ph/0009055, (2000); H. Zbinden, J. Brendel, N. Gisin, and W. Tittel, "Experimental test of non-local quantum correlation in relativistic configurations", Phys. Rev. A **63**, 022111 (2001).